\documentclass[%
reprint,
 amsmath,amssymb,
 aps,
]{revtex4-1}

\usepackage{graphicx}
\usepackage{dcolumn}
\usepackage{bm}
\usepackage{color}

\begin{document}

\preprint{APS/123-QED}

\title{Onsager's Variational Principle for the Dynamics of a Vesicle in a Poiseuille Flow}

\author{Yutaka Oya}
\author{Toshihiro Kawakatsu}%
\affiliation{%
 Department of Physics, Tohoku University, Sendai 980-8578, Japan
}%




\date{\today}

\begin{abstract}
We propose a systematic formulation of the migration behaviors of a vesicle in a Poiseuille flow based on Onsager's variational principle, which can be used to determine the most stable steady state.  
Our model is described by a combination of the phase field theory for the vesicle and the hydrodynamics for the flow field.  
The dynamics is governed by the bending elastic energy and the dissipation functional, the latter being  
composed of viscous dissipation of the flow field, dissipation of the bending energy of the vesicle, and the friction between the vesicle and the flow field.
We performed a series of simulations on 2-dimensional systems by changing the bending elasticity of the membrane, and observed 3 types of steady states, {\it i.e.} those with  slipper shape, bullet shape and snaking motion, 
and a quasi steady state with zig-zag motion.
We show that the transitions among these steady states can be quantitatively explained by evaluating the dissipation functional, which is determined by the competition between the friction on the vesicle surface 
and the viscous dissipation in the bulk flow.
\end{abstract}

\pacs{Valid PACS appear here}
\maketitle


%

\section{Introduction}

Migration of a bio-membrane in a narrow capillary is an important phenomenon for the understanding of the biological activities.  
For example, as was reported by Skalak {\it et al.}\cite{Skalak}, red blood cells circulating in blood vessels show a large variety of shapes, such as a symmetric parachute and an asymmetric slipper shapes.  
Especially, in the asymmetric shape, the vesicle shows tank-treading motion, where amphiphilic molecules rotate on the vesicle surface keeping the whole vesicle shape stationary.

Recently, to elucidate the physical origin of the behaviors of red blood cells in a capillary, several simulation methods have been developed.  
Using surface element method, Noguchi and Gompper simulated the shape transition of a vesicle between an oblate shape and a prolate shape\cite{Noguchi}.
Kaoui {\it et al.} showed with use of the boundary integral method that the shape transition of a vesicle between the parachute and the slipper shapes is triggered by a decrease in the velocity difference between the vesicle center and the unperturbed external Poiseuille flow\cite{Kaoui}.  
In these methods, the red blood cell is modeled as a closed surface which is discretized using a chain of bonds with fixed length (2 dimensional case in Ref.[3]) or using a triangular mesh (3 dimensional case in Ref.[2]).

Contrary to these studies, in the present work, we use field theories for both of the vesicle shapes and the flow field\cite{Biben,Biben_2,Q_Du}.  
By using the field theoretic description, we can easily evaluate the free energy and the dissipation functional of the system, which enables us to determine the equilibrium or steady states of the system quantitatively.
In the case of static simulations, the equilibrium shape of the vesicle is obtained by comparing the values of the total free energy of the vesicles among several candidates\cite{Seifelt}.  
Similarly to this free energy analysis in the static simulations, in the case of our dynamical simulations, 
stable steady states of the vesicle and the flow field are determined by comparing Onsager's dissipation functional instead of the free energy of the system\cite{Doi, Doi_2}. 

Although there are some candidates for the field theory for vesicles, such as self-consistent field theory{\cite{Wang,Lauw}} and Ota-Kawasaki theory{\cite{Uneyama,Uneyama_2}}, we choose the phase field theory(PFT) for describing the vesicle dynamics\cite{Biben,Biben_2,Q_Du,Campelo,Oya,Oya_2,Du}.  
Contrary to the other field theories, by using PFT, the interface between inside and outside regions of the vesicle can be described stably, and we can easily trace the moving interface driven by an external flow field.  
In this study, we extended the PFT to realize the vesicle dynamics governed by equation of motion based on the Onsager's variational principle.  
Especially, we applied this theory to the dynamics of a vesicle in a 2-dimensional Poiseuille flow in order to understand the origin of transitions of migration behaviors of the vesicle by analyzing the Onsager's dissipation functional.  
The paper is organized as follows.

In Section 2, we show the detail of our field theory based on Onsager's principle.  
In Section 3, we show our simulation method for the shape deformations of a vesicle in a 2-dimensional Poiseuille flow.  
In Section 4, we will show our simulation results.
Finally, in Section 5, we conclude this study.  

\section{Onsager's variational principle}

Onsager's principle is a fundamental framework that describes the complex non-equilibrium behaviors of soft-materials, such as liquid crystals, gels, polymers and colloids\cite{Doi, Doi_2}.  
Based on the Onsager's principle, the dynamical equations for the vesicle and the flow field can be derived.  
In this study, we consider a two component system composed of amphiphilic molecules that form the vesicle and a surrounding solvent.  
For simplicity, we assume that the mass density of the amphiphilic molecules and the solvent are the same.

First, we define the flow velocity field ${\bf v}({\bf r},{t})$ as
\begin{equation}
{\bf v}({\bf r},t) \equiv \phi_{\rm m}({\bf r},t) {\bf v}_{\rm m} ({\bf r},t) + (1- \phi_{\rm m}({\bf r},t)) {\bf v}_{\rm s}({\bf r},t), 
\end{equation}
where ${\bf v}_{\rm m}$ is the coarse-grained velocity field of the amphiphilic molecules that form the membrane of the vesicle, ${\bf v}_{\rm s}$ is the velocity of the solvent, and $\phi_{\rm m}$ is the volume fraction of the amphiphilic molecules, where we assumed an incompressibility condition on the total density of the membrane and the solvent.  
Among the three velocity fields ${\bf v},{\bf v}_{\rm m}$ and ${\bf v}_{\rm s}$, we choose ${\bf v}$ and ${\bf v}_{\rm m}$ as independent variables.  
Hereafter, unless explicitly stated, we will suppress the argument of time $t$ for simplicity. 
As $\phi_{\rm m}$ is a conserved variable, the following equation of continuity for $\phi_{\rm m}$ should be satisfied:
\begin{equation}
\label{cont}
\frac{\partial \phi_{\rm m}({\bf r})}{\partial t} = - \nabla \cdot \left( \phi_{\rm m}({\bf r}) {\bf v}_{\rm m}({\bf r}) \right).
\end{equation}
To obtain the equations of motion for ${\bf v}$, ${\bf v}_{\rm m}$ and $\phi_{\rm m}$, we introduce the Onsager's dissipation functional $R$ defined by
\begin{eqnarray}
\label{Onsager}
R \equiv \dot{F}_{\rm total} - \int p({\bf r}) \nabla \cdot {\bf v}({\bf r}) d{\bf r} + W_{\rm flow} + W_{\rm surface}.
\end{eqnarray}
The first term on the right-hand side of eq.(\ref{Onsager}) is the time derivative of the total free energy of the system $F_{\rm total}$, 
and the second term defines the Lagrangian multiplier $p({\bf r})$ for the incompressibility condition ($\nabla \cdot {\bf v}=0$), where $p({\bf r})$ corresponds to the local static pressure.  
The total free energy $F_{\rm total}$, and the third and the fourth terms on the right-hand side of eq.(\ref{Onsager}) are defined as
\begin{eqnarray}
\label{total_F}
F_{\rm total} \equiv F_{\rm bend} + \sigma \left( {\cal S} - {\cal S}_{0} \right) + \gamma \left( {\cal V} - {\cal V}_{0}  \right), \\
\label{W_flow}
W_{\rm flow} \equiv \frac{1}{4} \int \eta ({\bf r}) \left[ \nabla {\bf v}({\bf r}) + \left( \nabla {\bf v}({\bf r}) \right)^{T} \right]^{2} d{\bf r}, \\
\label{W_slip}
W_{\rm surface} \equiv \frac{1}{2} \int \frac{\phi_{\rm m}({\bf r})}{L({\bf r})} \left[ {\bf v}({\bf r}) - {\bf v}_{\rm m}({\bf r}) \right]^{2} d{\bf r}, 
\end{eqnarray} 
where $F_{\rm bend}$ is the Helfrich's bending energy\cite{Seifelt}, 
$\sigma$ and $\gamma$ are Lagrangian multipliers for the constraints on the total surface area ${\cal S}$ and on the total enclosed volume ${\cal V}$ that are fixed at the values ${\cal S}_{0}$ and ${\cal V}_{0}$, respectively.  
$\eta({\bf r})$ is the viscosity of the flow, and $L({\bf r})$ is the mobility of the amphiphilic molecules.  
In the integrand of the right-hand side of eq.(\ref{W_flow}), square of a second rank tensor ${\bf A}$ is defined by ${\bf A}^{2} \equiv \rm{trace}( {\bf A} \cdot {\bf A}  )$.
$W_{\rm flow}$ is the dissipation functional of the flow field and $W_{\rm surface}$ is the contribution from the friction between the vesicle and the flow field.
Here, we note that, in the limit of $L \rightarrow 0$, $W_{\rm surface}$ imposes the no-slip boundary condition on the solvent at the vesicle surface, while $W_{\rm surface}$ still has non-zero contribution.

Due to the requirement of the non-equilibrium thermodynamics, the time developments of the vesicle shape and the flow field are given by minimizing the Onsager's dissipation functional $R$, 
which leads to two conditions ${\delta R}/{\delta {\bf v}_{\rm m}({\bf r})} = {\bf 0}$ and ${\delta R}/{\delta {\bf v}({\bf r})} = {\bf 0}$.  
Therefore, we obtain the following equations of motion for ${\bf v}$ and ${\bf v}_{\rm m}$:
\begin{eqnarray}
\label{stokes}
\nabla \cdot \left[ \eta({\bf r}) \nabla {\bf v} ({\bf r}) \right] - \nabla p({\bf r}) - \phi_{\rm m}({\bf r}) \nabla \frac{\delta F_{\rm total}}{\delta \phi_{\rm m}} = {\bf 0}, \\
\label{vm}
{\bf v}_{\rm m}({\bf r}) = - L({\bf r}) \nabla \frac{\delta F_{\rm total}}{\delta \phi_{\rm m}} + {\bf v}({\bf r}).
\end{eqnarray} 
Equation (\ref{stokes}) represents the Navier-Stokes equation with Stokes approximation for a flow field with a low Reynolds number. 
The third term on the left-hand side of eq.(\ref{stokes}) corresponds to the stress due to the bending energy of the vesicle.  
By combining eqs.(\ref{cont}) and (\ref{vm}), we obtain the time evolution equation for the vesicle shape as
\begin{equation}
\label{time_phi}
\frac{\partial \phi_{\rm m}({\bf r})}{\partial t} = \nabla \cdot \left( L({\bf r}) \phi_{\rm m} \nabla \frac{\delta F_{\rm total}}{\delta \phi_{\rm m} ({\bf r}) } - \phi_{\rm m} ({\bf r}) {\bf v} ({\bf r})    \right), 
\end{equation}
where the first term on the right-hand side of eq.(\ref{time_phi}) represents the Fick's law of linear diffusion of the amphiphilic molecules, and the second term is the contribution from the advection due to the external flow field. 

\section{Simulation methods}
\subsection{Phase field equations}
In order to simulate the vesicle behaviors based on eq.(\ref{time_phi}), we use phase field theory(PFT)\cite{Biben,Biben_2,Q_Du,Campelo,Oya,Oya_2,Du}.  
In the PFT, the vesicle surface is represented by a scalar order parameter field $\psi({\bf r})$, which is called ``phase field'', where ${\bf r}$ is a position vector.  
Inside and outside regions of the vesicle are specified by positive and negative values of $\psi$, respectively.
Therefore the vesicle surface is defined by $\psi=0$.
As such an interface is a topological defect of the field $\psi$, the interface is stable and non-vanishing even if vesicle is migrated and deformed by the flow, which is an advantage of the PFT when simulating the flow behavior of the vesicle.

In the following, we rewrite eq.(\ref{time_phi}), {\it i.e.} equation of motion for the density field of the amphiphilic molecules, into a form of a time evolution equation for the phase field $\psi$.  
Using the Ginzburg-Landau model for binary mixtures, we define the distribution of the amphiphilic molecules of the vesicle as\cite{Oya,Oya_2}
\begin{equation}
\label{phi_2}
\phi_{\rm m}({\bf r}) = \frac{1}{2} \left( 1 - \psi({\bf r})^{2}  \right)^{2} + \epsilon^{2} | \nabla \psi({\bf r}) |^{2}, 
\end{equation}   
where $\epsilon$ is the interface thickness.  
It is easy to check that $\phi_{\rm m}({\bf r})$ defined by eq.(\ref{phi_2}) takes non-zero value only on the membrane surface.
As is well-known in the differential geometry\cite{Seifelt},
the mean curvature of the vesicle surface is obtained by a variation of the total surface area $S$, which is proportional to the total component of the amphiphilic molecules
$\int \phi_{\rm m} d{\bf r}$ with respect to an infinitesimal displacement of the vesicle surface in its perpendicular direction.  
Therefore, we obtain the following formula\cite{Du}
\begin{equation}
\label{mean} 
H({\bf r}) = - \psi({\bf r}) + \psi({\bf r})^{3} - \epsilon^{2} \nabla^{2} \psi({\bf r}),
\end{equation}
where $H({\bf r})$ is a mean curvature multiplied by a constant that takes account of the dimensionality.  
Using eq.(\ref{mean}), the Helfrich's bending energy for vesicle shapes are obtained by
\begin{equation}
\label{bend}
F_{\rm bend} = \frac{\kappa}{2 \epsilon^{3}} \int H({\bf r})^{2} d{\bf r},
\end{equation} 
where $\kappa$ is the bending modulus of the vesicle.\\  
Now, we rewrite the left-hand side of eq.(\ref{time_phi}) by using the following identity
\begin{equation}
\label{new_1}
\frac{\partial \phi_{\rm m}({\bf r})}{\partial t} 
= \int \frac{\partial \psi({\bf r}')}{\partial t} \frac{\delta \phi_{\rm m}({\bf r})}{\delta \psi({\bf r}')} d{\bf r}'.
\end{equation} 
Substituting eq.(\ref{phi_2}) into eq.(\ref{new_1}), we obtain the following relation
\begin{eqnarray}
\label{psi_time_1}
\frac{\partial \phi_{\rm m}({\bf r})}{\partial t} = 2 \epsilon^{2} \nabla \cdot \left( \frac{\partial \psi({\bf r})}{\partial t} \nabla \psi({\bf r})   \right) + 2H({\bf r}) \frac{\partial \psi({\bf r})}{\partial t}. \nonumber \\
\end{eqnarray} 
We first solve eq.(\ref{psi_time_1}) for a flat membrane with $H=0$.  
Using the Green function of Coulomb potential generated by a point charge,  
we obtain 0-th order solution of eq.(\ref{psi_time_1}), {\it i.e.} a flat membrane solution, as follows:
\begin{equation}
\label{psi_time_2}
\left( \frac{\partial \psi({\bf r})}{\partial t}  \right)_{0} = \frac{1}{2 \epsilon^{2}} \frac{\nabla \psi({\bf r})}{ |\nabla \psi({\bf r}) |^{2} } \cdot \int {\bf G}({\bf r}-{\bf r}') \frac{\partial \phi_{\rm m}({\bf r}')}{\partial t}d{\bf r}',
\end{equation}
where ${\bf G}$ is Green function that satisfies $\nabla^{2} {\bf G} ({\bf r}-{\bf r}')= \delta ({\bf r}-{\bf r}')$.  
Substituting eq.(\ref{psi_time_2}) into the second term on the right-hand side of eq.(\ref{psi_time_1}) and repeating the same procedure that was done for the 0-th order solution, 
we obtain the time evolution equation for the phase field up to the first order in the mean curvature $H$ as
\begin{eqnarray}
\label{new_2}
\frac{\partial \psi({\bf r})}{\partial t} = \frac{1}{2 \epsilon^{2}} \frac{\nabla \psi({\bf r})}{\mid \nabla \psi({\bf r}) \mid^{2} }
\cdot \int {\bf G}({\bf r}-{\bf r}') \Biggl\{ \frac{\partial \phi_{\rm m}({\bf r}')}{\partial t}  \nonumber \\
- 2 H({\bf r}') \left( \frac{\partial \psi({\bf r}')}{\partial t} \right)_{0} \Biggr\} d{\bf r}' + o(H^{2}),  
\end{eqnarray}
where $o(H^2)$ means terms of order $H^{2}$ and higher orders.
By using the identity
\begin{equation}
\frac{\delta F_{\rm total}}{\delta \psi({\bf r})} =\int \frac{\delta F_{\rm total}}{\delta \phi_{\rm m}({\bf r}')} \frac{\delta \phi_{\rm m}({\bf r}')}{\delta \psi({\bf r}) } d{\bf r}',
\end{equation}
the functional derivative of the total free energy with respect to the membrane shape ${\delta F_{\rm total}}/{\delta \phi_{\rm m}}$ 
is also obtained with the same procedure as was used in the derivation of eq.(\ref{new_2}).
The resultant formula leads to
\begin{equation}
\label{new_3}
\frac{\delta F_{\rm total}}{\delta \phi_{\rm m}({\bf r})} 
= - \frac{1}{2 \epsilon^{2}} \int  {\bf G}({\bf r}-{\bf r}') \cdot \frac{\nabla' \psi({\bf r}')}{\mid \nabla \psi({\bf r}') \mid^{2}}
\frac{\delta F_{\rm total}}{\delta \psi({\bf r}')} d{\bf r}' + o(H^{2}).
\end{equation}
Equation(\ref{new_3}) is used in the Navier-Stokes equation eq.(\ref{stokes}) as the stress due to the bending energy of the membrane.

As the simplest example of eq.(\ref{new_2}), we first consider the case of no advection, {\it i.e.} the case where the second term on the right-hand side of eq.(\ref{time_phi}) is dropped.  
Using eqs.(\ref{time_phi}) and (\ref{psi_time_1}), we obtain the following equation of motion for $\psi({\bf r})$(detail is shown in {\it Appendix}):
\begin{equation}
\label{noncons}
\frac{\partial \psi({\bf r})}{\partial t} = - \frac{L({\bf r})}{2 \epsilon^{2}} \frac{\delta F_{\rm total}}{\delta \psi({\bf r})} + o(H^{2}).
\end{equation}
Equation(\ref{noncons}) means that the conservation equation for the density distribution of the amphiphilic molecules $\phi_{\rm m}$ reduces to 
a non-conserved dynamical equation for the phase field $\psi$ up to the leading order in the mean curvature $H$.  

Using eq.(\ref{noncons}), the final form of the time evolution equation for $\psi$ with the advection term is given by 
\begin{eqnarray}
\label{psi_time_3}
\frac{\partial \psi({\bf r})}{\partial t} = -\frac{L({\bf r})}{2 \epsilon^{2}} \frac{\delta F_{\rm total}}{\delta \psi({\bf r})}
- \frac{1}{2\epsilon^{2}} \frac{\nabla \psi({\bf r})}{|\nabla \psi({\bf r})|^{2}} \cdot \Biggl\{   \phi_{\rm m}({\bf r}) {\bf v}({\bf r})    \nonumber \\
+ \int d{\bf r}' {\bf G}({\bf r}-{\bf r}') H({\bf r}')  \nabla' \psi({\bf r}') \cdot {\bf v}({\bf r}')  \Biggr\} 
+ o(H^{2}). \nonumber \\
\end{eqnarray}
Equation(\ref{psi_time_3}) and eq.(\ref{stokes}) where eq.(\ref{new_3}) is substituted into the 3rd term on the left hand side form a closed set of time evolution equations for our simulations. 
For simplicity, hereafter, we assume that the system is 2-dimensional, and $L({\bf r})$ and $\eta({\bf r})$ are constant values $L$ and $\eta$.   
We also assume a 2-dimensional Poiseuille flow, where we introduce $x$ and $y$ coordinates in the parallel and perpendicular directions to the unperturbed flow.  
We solve the Navier-Stokes equation eq.(\ref{stokes}) under the incompressibility condition with periodic boundary condition in the $x$-direction at $x=0$ and $x=x_{\rm max}$ and non-slip boundary condition at $y=0$ and $y=y_{\rm max}$, where $x_{\rm max}$ and $y_{\rm max}$ are the length and the width of the flow channel.
%
\subsection{Non-dimensional parameters}
Our model system is characterized by several non-dimensional parameters.  
In our 2-dimensional model, ${\cal V}$ and ${\cal S}$ in eq.(\ref{total_F}) represent the total enclosed area and the perimeter length of the vesicle, respectively. 
We define the vesicle size $R_{0}$ as the radius of a perfect circle that has the same perimeter length ${\cal S}$ of the vesicle.  
With use of the maximum velocity of the unperturbed Poiseuille flow $V_{\rm max}$({\it i.e.} the velocity at the center of the channel), 
the bending modulus of the vesicle $\kappa$ and the density of the solvent $\rho$, 
we introduce the 3 non-dimensional parameters that characterize the system, {\it i.e.} the reduced volume $\upsilon$, the Reynolds number ${\rm Re}$ and the capillary number ${\rm Ca}$ defined as
$\upsilon = 4 \pi {\cal V}/  {\cal S}^{2} $, 
${\rm Re} =  2 \rho V_{\rm max} R_{0} / \eta$ and 
${\rm Ca} =  {4 \eta V_{\rm max} R_{0}^{3}}/ \kappa y_{\rm max}$.
Here, ${\rm Re}$ is a measure of the ratio between the viscous force and the inertial force
and ${\rm Ca}$ is the ratio between the viscous force and the stress force, the latter originating from the bending elasticity of the membrane.  
In the case of a red blood cell in a blood vessel, the Reynolds number and the capillary number usually satisfy ${\rm Re}<0.01$ and ${\rm Ca} < 0.1$, respectively.

We performed 2-dimensional simulations using eqs.(\ref{stokes}) and (\ref{psi_time_3}).  In eq.(\ref{stokes}), 
for the stability of the numerical scheme, 
we relaxed the Stokes approximation by introducing the inertia term that was neglected in eq.(\ref{stokes}).  We confirmed that this addition does not have any appreciable contributions to the dynamics of the whole system 
except for rapid oscillations of the quantities around their average value.  
As long as we are concerned with the slow motion of the flow and the vesicle, these rapid oscillations can be averaged out.

It is convenient to introduce two alternating non-dimensional control parameters for the numerical scheme instead of the physical parameters $\rm{Ca}$ and $\rm{Re}$ defined above.  In order to define these control parameters, we rewrite our original equations, eqs.(\ref{stokes}) and (\ref{time_phi}), in non-dimensional forms.
Using the density of the fluid $\rho$ and an arbitrary chosen constant $\alpha$ which corresponds to an apparent Reynolds number, we set the unit length $x_{0}$, the unit mass $m_{0}$ and the unit time $t_{0}$ as
$x_{0}=\left( \alpha \rho L \kappa/ \eta \right)^{\frac{1}{3}}$, $m_{0}=\alpha \rho^{2} L \kappa / \eta $ and $t_{0}= \left( \alpha \rho L \kappa / \eta  \right)^{\frac{5}{3}} / L \kappa $, respectively. 
Using these units, we rewrite eqs.~(\ref{stokes}) and (\ref{time_phi}) into non-dimensional forms.
%
Such a non-dimensionalization tells us that our model equations have a characteristic non-dimensional parameter $ ( 1/{\rm Ca} )_{0} = \left( \alpha^{5} \rho^{2} \kappa^{2} / L \eta^{5}  \right)^{\frac{1}{3}}$ and a non-dimensional scaled free energy functional $ \bar{F}_{\rm total}$ defined by $\bar{F}_{\rm total} = F_{\rm total}/\kappa$.

Hereafter, we treat all quantities as non-dimensionalized using these units $x_{0}$, $m_{0}$ and $t_{0}$.
In the above derivation of the non-dimensionalized forms of eqs.(\ref{stokes}) and (\ref{time_phi}), 
the units used for the nondimensionalization were not chosen according to the physical properties of the flow.  
Such flow characteristics are specified by $\rm Re$ and $\rm Ca$, which are now rewritten as 
\begin{eqnarray}
\label{rey}
{\rm Re} = {2 \rho V_{\rm max} R_{0}} {\Big/} \alpha , \\
\label{cap}  
{\rm Ca} = \frac{ 4 \alpha V_{\rm max} R_{0}^{3}}{ y_{\rm max}}{\Big/} \left( \frac{1}{\rm Ca} \right)_{0}.   
\end{eqnarray}
In general, the flow field can be well described by the Navier-Stokes equation with the Stokes approximation for systems with low Reynolds number that satisfies ${\rm Re}< 0.1$.  This condition corresponds to a large value of $\alpha$.  
In our simulations, we choose $\alpha=25$ while changing $V_{\rm max}$, $(1/{\rm Ca})_{0}$ and $\upsilon$.  
This choice of $\alpha$ validates the use of the Stokes approximation because ${\rm Re}$ defined by eq.(\ref{rey}) satisfies ${\rm Re}<0.1$.  

In our simulation, we used non-dimensionalized forms of eqs.(\ref{stokes}) and (\ref{psi_time_3}) to solve the dynamic motions of the vesicle and the flow field, respectively.
In order to solve these equations, 
we used a 2-dimensional mesh that has $128 \times 128$ mesh points with a mesh width $\Delta x = 0.5$, time step width $\Delta t = 0.001$ and membrane thickness $\epsilon = 1.0$.

As the initial state of the simulations, the vesicle is placed at $Y_{\rm G}=-5$, where $Y_{\rm G}$ is the lateral position of the center of mass of the vesicle
in the vertical direction to the flow velocity.  
We confirmed that the final steady states of the vesicle are independent of this initial position $Y_{\rm G}$ as long as $Y_{\rm G}$ is small compared to the channel width $y_{\rm max}=32.0$.  
It is known that this ergodicity property is not always satisfied, for example for a vesicle that contains a highly viscous solvent\cite{Farutin}.

\section{Results and discussions}

\subsection{Various steady states}

\begin{figure}[t]
   \begin{center}
   \includegraphics[width=60mm]{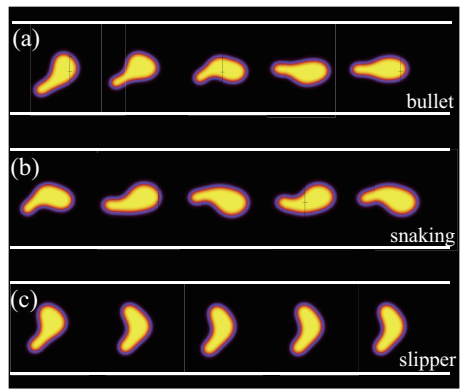}
\caption{\label{v07_shapes} 
The time evolutions of the shapes of the vesicle with $\upsilon=0.7$, $V_{\rm max}=0.05$, and 
$(1/{\rm Ca})_{0}$ equals to (a)1, (b)20 and (c)25, respectively. 
In the steady states, (a) and (c) represent bullet and slipper shapes.  
We also observe a snaking oscillation in (b).}
  \end{center}
  \vspace{-7mm}
\end{figure}
\begin{figure}[t]
   \begin{center}
   \includegraphics[width=80mm]{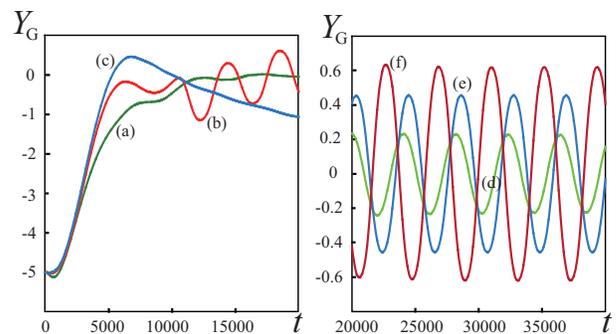}
\caption{\label{YG_v07_Vmax500_Ca0} 
The time evolutions of the lateral positions of the center of mass of the vesicles with $\upsilon=0.7$ and $V_{\rm max}=0.05$ for different value of $(1/{\rm Ca})_{0}$. 
Left-hand side figure shows the lateral positions of the same vesicles as those shown in fig.~\ref{v07_shapes} in the early stage ({\it i.e.} (a)$(1/{\rm Ca})_{0}=1$,
(b)20 and (c)25, respectively).  
Right-hand side figure shows the lateral positions of the vesicle 
in the late stage of the snaking motion for the cases with (d)$(1/{\rm Ca})_{0}=10$, (e)15 and (f)20, respectively. }
  \end{center}
\end{figure}
In fig.\ref{v07_shapes}, we show typical behaviors of the vesicle with $\upsilon=0.7$ and $V_{\rm max}=0.05$. 
In this case, we found three steady states of the vesicle, {\it i. e.} a bullet shape, a slipper shape and a snaking oscillation.
The bullet shape is a symmetric shape, where the vesicle shape is elongated in the flow direction(fig.\ref{v07_shapes}(a)).  
On the other hand, the slipper shape is an asymmetric shape, where the direction of the elongation axis of the vesicle is perpendicular to the flow velocity(fig.\ref{v07_shapes}(c)).  These bullet and slipper shapes are known to be typical steady states of the vesicle in a Poiseuille flow\cite{Noguchi,Kaoui}.
On the other hand, the snaking oscillation shown in fig.\ref{v07_shapes}(b) was rather recently discovered \cite{Kaoui_2}.  
In the snaking oscillation, the vesicle shape is elongated in a similar manner as the bullet shape. 
However, the rear end of the vesicle is temporarily bent just like the shape of a snake in its locomotion. 

In fig.\ref{YG_v07_Vmax500_Ca0}, we show the lateral positions of the center of mass of the vesicle $Y_{\rm G}$ in the vertical direction to the flow velocity. 
In these figures, $(1/{\rm Ca})_{0}$ is changed, while the maximum value of the flow velocity $V_{\rm max}$ is kept constant at the same value as was used in fig.~\ref{v07_shapes}, {\it i.e.} $V_{\rm max}=0.05$. 
The left-hand side figure shows the lateral positions of the center of mass of the vesicle, where $(1/{\rm Ca})_{0}$ equals to 1, 20 and 25, respectively. 
When $(1/{\rm Ca})_{0}$ is increased, the stationary shape of the vesicle changes from (a)the bullet shape to (c)the slipper shape.    
The snaking behavior is found for (b)the intermediate value. 
In the right-hand side figure of fig.\ref{YG_v07_Vmax500_Ca0}, we show the temporal behaviors of the lateral positions of the center of mass of the vesicle for the snaking behavior. 
With increasing $(1/{\rm Ca})_{0}$, which corresponds to increasing the bending modulus $\kappa$, the amplitude of the oscillation of the lateral position is increased, 
which means that the stress of the flow field induced by the bending elasticity of the vesicle is important for the snaking behaviors.  
\subsection{Quantitative determination of stable steady state using Onsager's dissipation functional}

  \begin{figure}[t]
   \begin{center}
   \includegraphics[width=80mm]{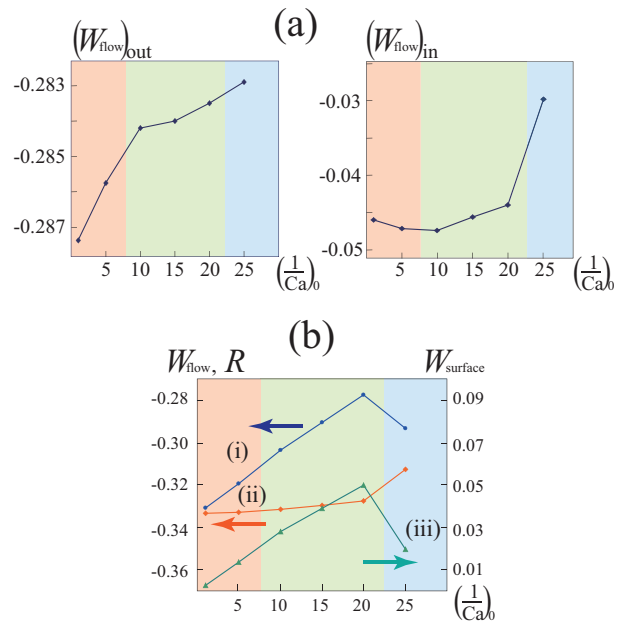}
\caption{\label{ED}
Components of the dissipation functional for the case with $V_{\rm max}=0.05$ and $\upsilon = 0.7$ are shown.  
(a)Dissipation functionals of the external(left) and the internal(right) fluids of the vesicles, where sum of these components corresponds to $W_{\rm flow}$ shown in (b).  
(b)Three curves express 
(i)the total dissipation $R$ $\left( {\rm eq.}(\ref{Onsager}) \right)$, (ii)dissipation due to the flow field $W_{\rm flow}$ $\left( {\rm eq}.(\ref{W_flow}) \right)$ and (iii)dissipation due to surface friction $W_{\rm surface}$ $\left( {\rm eq.}(\ref{W_slip}) \right)$, respectively.  
Vertical axis on the left edge of each figure represents the values of $R$ and $W_{\rm flow}$, while the right edge represents $W_{\rm surface}$.
The horizontal axes of (a) and (b) represent $(1/{\rm Ca})_{0}$.
In these figures, the vesicle shows bullet shapes (orange regions:$(1/{\rm Ca})_{0}<5$), snaking motions (green regions:$10<(1/{\rm Ca})_{0}<20$) and slipper shapes (blue regions:$(1/{\rm Ca})_{0}>25$). 
}
	\end{center}
   \end{figure}
To quantitatively explain the transitions among different steady states of the vesicle shown in the previous subsection, 
we focus on the dissipation functional defined by eqs.(\ref{Onsager})-(\ref{W_slip}) (for simplicity, we call this just ``dissipation'').  
In the steady states, contribution from the time derivative of the total free energy $\dot{F}_{\rm total}$ is vanishing. 
(In the case of the snaking motions, time average of $\dot{F}_{\rm total}$ is zero instead of the instantaneous value of $\dot{F}_{\rm total}$.)  
Moreover, we found that the second terms on the right-hand side of eq.(\ref{Onsager}) is negligible due to the incompressibility conditions of the flow field.  
Figure \ref{ED} shows the dependences of the components of the dissipation on $(1/{\rm Ca})_{0}$, where the value of $(1/{\rm Ca})_{0}$ is a measure of the importance of the bending elasticity of the vesicle compared with that of the fluid. 

Figure \ref{ED}(a) shows the contributions from the fluid flows in the outside and inside regions of the vesicle.  
These values represent the deviations in the dissipation from that in the unperturbed Poiseuille flow.
When the value of $(1/{\rm Ca})_{0}$ is increased, bullet, snaking and slipper shapes appear in this order. At each transition point between different shapes, we observe a kink in the curve, which indicates an occurrence of a transition from one stable branch to another.  This is analogous to the 1st order phase transition.

In the bullet region ($(1/{\rm Ca})_{0} < 5$), the dissipation of the outside fluid is increasing as $(1/{\rm Ca})_{0}$ is increased while the change in the dissipation of the inside fluid is less pronounced.  
The extra dissipation of the outside fluid from the unperturbed Poiseuille flow is due to the disturbance near the surface of the vesicle, which plays the role of an essentially stick boundary for the fluid.  
When $(1/{\rm Ca})_{0}$ is small (a flexible vesicle), the vesicle shape is deformed by the fluid flow so that the dissipation of the flow is minimized.  
Such an adjustment of the vesicle shape becomes less effective when $(1/{\rm Ca})_{0}$ becomes large because of the increasing stiffness of the vesicle.  
This leads to an increase in the dissipation of the outside fluid.  
On the other hand, the inside fluid is static in the frame of the enclosing vesicle because the bullet shape does not change its shape with no flow on the membrane just like a rigid container for the inside fluid.  
In fig.\ref{ED}(b)(ii), the sum of the contributions from the inside and the outside fluids is shown.
According to the above discussion on fig.\ref{ED}(a), the slight increase of this sum in the bullet region is coming from the outside fluid.  
On the other hand, fig.\ref{ED}(b)(iii) shows the dissipation due to the friction between the membrane and the fluid(both inside and outside of the vesicle) along the vesicle surface.  
One can observe that the increase of $(1/{\rm Ca})_{0}$ induces an increase of the dissipation due to the friction at the vesicle surface, 
which is much larger than the increase in the viscous dissipation of the bulk fluid.  
Thus, the main contribution to the dissipation of the whole system in the bullet region is the viscous dissipation of the outside fluid at the vesicle surface(hereafter we call this ``surface friction'').

In the snaking region($10 < (1/{\rm Ca})_{0} < 20$), the slope of the curve for the outside fluid (left-side figure of fig.\ref{ED}(a)) is smaller than that in the bullet region.  
This behavior means that the perturbation to the outside fluid imposed by the vesicle is more relaxed for the snaking motion than the bullet shape.  
As there is a tank-treading motion of the membrane for the snaking motion as well as the change in its whole shape, 
it is easier for the snaking vesicle to adjust itself to the outside fluid than the bullet shape so that the friction of the outside fluid is reduced.  
This adjustment is realized at the sacrifice of the increase in the dissipation of the inside fluid, which is almost the same amount as that of the outside fluid.  
The sum of these two contributions gives the increase in fig.\ref{ED}(b)(ii) in the snaking region, which has a slightly larger slope than that in the bullet region.  
This increase is compensated by the reduction in the slope of the dissipation due to the surface friction (fig.\ref{ED}(b)(iii)) induced by the tank-treading motion of the vesicle surface.  
As a result, the total dissipation $R \left( {\rm fig.\ref{ED}(b)(ii)} \right)$ has a smaller slope than that in the bullet region.  
The difference in the slopes of the total dissipation between bullet and snaking regions leads to a transition from the bullet branch to the snaking branch at around $(1/{\rm Ca})_{0} \sim 5$.

For the slipper region($25 < (1/{\rm Ca})_{0}$), the effect of the tank-treading motion is much more pronounced, 
which is indicated by the abrupt decrease in the dissipation due to the surface friction (fig.\ref{ED}(b)(iii)).  
Such a decrease in the dissipation due to the surface friction overcomes the increase in the dissipation due to the fluid(fig.\ref{ED}(b)(ii)), 
and the total dissipation $R \left( {\rm fig.\ref{ED}(b)(i)} \right)$ has a smaller slope than that in the snaking region.  
This is again the origin of the second transition from the snaking shape to the slipper shape at around$(1/{\rm Ca})_{0} \sim 22$.
\begin{figure}[t]
   \begin{center}
   \includegraphics[width=80mm]{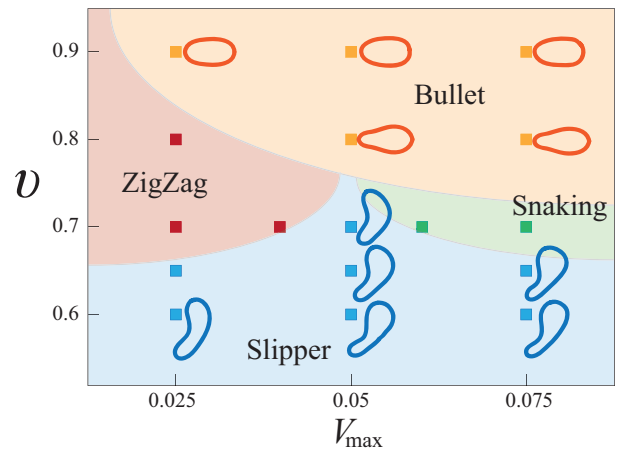}
\caption{\label{phase_diagram_1} 
Phase diagram of the behaviors of the vesicle with $(1/{\rm Ca})_{0}=25$. 
Horizontal and vertical axes mean $V_{\rm max}$ and $\upsilon$.
Each symbol represents the simulated point.  
Blue symbol represents the slipper shape, orange one the bullets shape, 
red one the zigzag motion and the green one the snaking motion, respectively.  
The stationary slipper and bullet shapes are shown besides the symbols.  
The boundaries are guide to the eyes. 
}
  \end{center}
    \vspace{-3mm}
\end{figure} 
\subsection{Steady state phase diagram}
In fig.\ref{phase_diagram_1}, we show the phase diagram of the behaviors of the vesicles with $(1/{\rm Ca})_{0}=25$.
The horizontal and the vertical axes represent $V_{\rm max}$ and $\upsilon$, respectively. 
When the reduced volume $\upsilon$ becomes larger, the vesicle changes its shape from an asymmetric shape(slipper) to a symmetric shape(bullet).  
Although this symmetric-asymmetric transition is qualitatively similar to that reported in preceding researches\cite{Kaoui,Kaoui_2,Shi}, there is a certain
difference between our results and the results reported by Kaoui {\it et al.}\cite{Kaoui_2}.  In order to explain this difference, let us focus on the steady shapes of the vesicle in the case $\upsilon=0.6$, for which we obtain only slipper shapes as is shown in fig.\ref{phase_diagram_1}.  
In this case, the capillary number ${\rm Ca}$ and the measure of the confinement of the vesicle inside the capillary $2R_{0}/y_{\rm max}$ take the values 
${\rm Ca}> 4.5$ and $2R_{0}/y_{\rm max}\simeq 0.44$, respectively.  
For the same conditions, however, Kaoui {\it et al.} showed that 
the slipper shape and the parachute shape are the steady shapes instead of the slipper shape we obtained\cite{supple}.  
The reason for this difference between our result and their result is the dissipation due to the friction at the membrane surface $W_{\rm surface}$ which is included only in our model.  
By combining eqs.(\ref{W_slip}) and (\ref{vm}), we obtain the relation 
\begin{equation}
\label{W_slip_2}
W_{\rm surface}= \left( \frac{1}{\rm Ca} \right)_{0} \int \frac{ \phi_{\rm m}}{2} \left\vert \nabla \frac{\delta \bar{F}_{\rm total}}{\delta \phi_{\rm m}} \right\vert^{2} d{\bf r},
\end{equation}
where $| \nabla \left( {\delta \bar{F}_{\rm total}}/{\delta \phi_{\rm m}} \right) |$ corresponds to 
the diffusion flux divided by the mobility $L$ of the amphiphilic molecules on the vesicle surface
that tend to reduce the total free energy of the vesicle $\bar{F}_{\rm total}$.  
The occurrence of such a diffusion motion of the amphiphilic molecules on the vesicle surface is the
reason why the vesicle with a large deformation is not preferable
with respect to $W_{\rm surface}$.  
Generally speaking, the slipper-shaped vesicle shows a smaller deformation 
than the parachute-shaped vesicle, because the amphiphilic molecules 
on the slipper-shaped vesicle undergo a tank-treading motion,
which prevents large deformation of the vesicle due to the shear force of the external flow. 
As a result, the slipper shape is selected as the stable state.

\begin{figure}[t]
   \begin{center}
   \includegraphics[width=80mm]{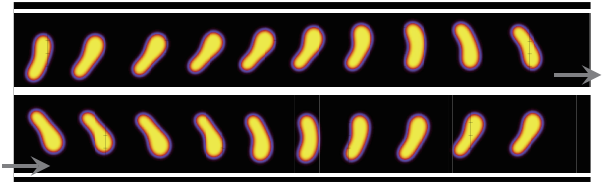}
\caption{\label{zigzag} 
The zig-zag motion of the vesicle with $\upsilon=0.7$, $V_{\rm max}=0.025$ and $(1/{\rm Ca})_{0}=25$.}
  \end{center}
\end{figure} 

In fig.\ref{phase_diagram_1}, we found two types of (damping) oscillations of the vesicle motions, {\it i. e.} snaking and zig-zag motions, which can be observed in the region between the stable regions of the slipper and bullet shapes.
The zig-zag motion is a damping oscillation in which the vesicle takes the parachute and the slipper shapes alternatively as is shown in fig.\ref{zigzag}. 
The vesicle moves toward the wall of the capillary with the slipper shape and then turns.
The parachute shape appears at this turning point.
After such a turning, the vesicle moves toward the other wall again with the slipper shape. 
It is difficult to obtain the final stationary shape of the vesicle in this case
because the period of the damping oscillation is very long. 
Therefore, we cannot identify which shape is chosen as the stationary shape in the region of zig-zag oscillations in Fig.\ref{phase_diagram_1}.

So far, we have shown two types of oscillations, {\it i. e.} the snaking motion and the zig-zag motion.  
In addition to these long-lived oscillations, we have also found a damping vacillating-breathing oscillation\cite{Misbah} in the region $\upsilon < 0.65$.
These oscillations have been reported in preceding works, where the vesicle is modeled with discretized meshes in 2-dimensions\cite{Shi, Shi_2} and 3-dimensions\cite{Noguchi_2, Fedosov}.
In these works, the vesicle deformations are induced by two different forces, {\it i.e.} the force due to the flow field and that due to the bending elasticity of the vesicle, both of which are also taken into account in our simulations in the diffusion of the amphiphilic molecules. 
On the other hand, Kaoui {\it et al.} proposed another model where the vesicle deforms due to the flow field without any friction.\cite{Kaoui_2}  
In this case, they reported only snaking oscillations and they did not observe the zig-zag and the vacillating-breathing oscillations. 
This result means that the various temporal oscillations are more pronounced by the contribution due to the friction at the membrane surface $W_{\rm surface}$ to the dissipation.
Because, in the model that includes both forces due to the flow field and the bending elasticity, the dissipation of flow field $W_{\rm flow}$ can compete not only with $\dot{F}_{\rm total}$ but also with $W_{\rm surface}$, 
and this competition is essential for the oscillation behaviors of the vesicle.

\section{Conclusion}
In this article, by using a combination of the PFT and Navier-Stokes equation, we derived, up to the first order in the local mean curvature of the vesicle, a set of dynamical equations which are equivalent to those obtained using the Onsager's principle.  Using this model, we performed dynamical simulations of the vesicle in a 2-dimensional Poiseuille flow which is a model of a red blood cell in a blood vessel.  
We simulated this model and found several types of behaviors of the vesicle, such as slipper shape, bullet shape, snaking motion, zig-zag motion and vacillating-breathing motion.  
An analysis based on the Onsaer's dissipation functional revealed that the origin of these behaviors is the friction between the vesicle surface and the flow field of the solvent.
Moreover, we should note the main differences between 2 and 3 dimensional systems discussed in ref.\cite{Fedosov}, {\it i.e.} (i) the absence of the slipper shape in 3-dimensional case and (ii) the absence of the tumbling behavior in the 2-dimensional case with a small viscosity contrast between outer and inner regions of the vesicle.  Our results do not contradict with these findings, and can give a reason for the stability of each of the stable behaviors.

\begin{center}
{\bf Acknowledgements}
\end{center}
The authors thank M.Imai, T.Taniguchi, T.Murashima and Y.Sakuma for valuable discussions.
The present work is supported by the Grant-in-Aid for Scientific Research from the Ministry of Education, Culture, Sports, Science, and Technology of Japan, and by the Global-COE program at Tohoku University.
  \vspace{-1mm}

\section*{Appendix}

In this Appendix, we derive eq.(\ref{noncons}), {\it i.e.} the solution of eq.(\ref{psi_time_1}), that is correct up to the 1st order in $H({\bf r})$.  
Substituting eq.(\ref{time_phi}) into the left-hand side of eq.(\ref{psi_time_1}), we obtain the following relation:
\begin{eqnarray}
\label{13_1}
\nabla \cdot \left( L({\bf r}) \phi_{\rm m}({\bf r}) \nabla \frac{\delta F_{\rm total}}{\delta \phi_{\rm m} ({\bf r})} 
- \phi_{\rm m}({\bf r}) {\bf v} ({\bf r}) \right) \nonumber \\
= 2 \epsilon^{2} \nabla \cdot \left( \frac{\partial \psi({\bf r})}{\partial t} \nabla \psi ({\bf r})    \right)
+2 H({\bf r}) \frac{\partial \psi({\bf r})}{\partial t}. 
\end{eqnarray}
In order to mimic the essence of the derivation, we consider the case without advection (i.e. ${\bf v}({\bf r})=0$ ).  
Then, eq.(\ref{13_1}) is simplified as
\begin{eqnarray}
\label{13_2}
\nabla \cdot \left( L({\bf r}) \phi_{\rm m}({\bf r}) \nabla \frac{\delta F_{\rm total}}{\delta \phi_{\rm m} ({\bf r})} \right)
&=& 2 \epsilon^{2} \nabla \cdot \left( \frac{\partial \psi({\bf r})}{\partial t} \nabla \psi ({\bf r})    \right)   \nonumber \\
&~& + 2 H({\bf r}) \frac{\partial \psi({\bf r})}{\partial t}.
\end{eqnarray}
As $\psi ({\bf r})$ is driven by the curvature of the vesicle surface, 
both of $\left( \partial \psi({\bf r})/\partial t \right)$ and $\left( \delta F_{\rm total} / \delta \phi_{\rm m}({\bf r}) \right)$ 
are first order in $H({\bf r})$.  
Then, eq.(\ref{13_2}) leads to  
\begin{equation}
\label{13_3}
\nabla \cdot \left( L({\bf r}) \phi_{\rm m}({\bf r}) \nabla \frac{\delta F_{\rm total}}{\delta \phi_{\rm m} ({\bf r})} \right)
= 2 \epsilon^{2} \nabla \cdot \left( \frac{\partial \psi({\bf r})}{\partial t} \nabla \psi ({\bf r})    \right)
+o(H^{2}).
\end{equation}
By integrating both sides of eq.(\ref{13_3}), we obtain
\begin{equation}
\label{13_4}
L({\bf r}) \phi_{\rm m} ({\bf r}) \nabla \frac{\delta F_{\rm total}}{\delta \phi_{\rm m}} 
=2 \epsilon^{2} \frac{\partial \psi({\bf r})}{\partial t} \nabla \psi({\bf r}) + {\bf A}({\bf r}) + o(H^{2}),
\end{equation}  
where ${\bf A}$ is an arbitrary vector field that satisfies the relation $\nabla \cdot {\bf A}=0$.  
As this vector field ${\bf A}$ does not affect the dynamics of the system, we can drop it without loss of generality.  
Multiplying $\nabla \psi ({\bf r})$ to both sides of eq.(\ref{13_4}) and dividing the resultant equation with $2 \epsilon^{2} | \nabla \psi ({\bf r}) |^{2} $, we obtain the expression for $\partial \psi({\bf r}) / \partial t$ up to the 1st order in $H$ as:
\begin{equation}
\label{13_5}
\frac{\partial \psi({\bf r})}{\partial t} = \frac{L({\bf r})}{2 \epsilon^{2}} \phi_{\rm m} ({\bf r}) \frac{\nabla \psi({\bf r})}{\mid \nabla \psi({\bf r}) \mid^{2} } \cdot \nabla \frac{\delta F_{\rm total}}{\delta \phi_{\rm m}({\bf r})} + o(H^{2}).
\end{equation}
We can further simplify this equation by using the following relation for the equilibrium profile of $\psi({\bf r})$
\begin{eqnarray}
\label{13_6}
\phi_{\rm m}({\bf r}) &=& \frac{1}{2} \left( 1 -\psi({\bf r})^{2}  \right)^{2} + \epsilon^{2} \mid \nabla \psi({\bf r}) \mid^{2} \nonumber \\
                      &=& 2 \epsilon^{2} \mid \nabla \psi({\bf r}) \mid^{2} + o(H), 
\end{eqnarray}
where we used the expression for the equilibrium profile $\psi({\bf r})= \tanh \left( {\bf r}/ \sqrt{2} \epsilon \right) + o(H)$.  
Substituting eq.(\ref{13_6}) into eq.(\ref{13_5}), we obtain  
\begin{equation}
\label{13_7}
\frac{\partial \psi({\bf r})}{\partial t} = L({\bf r}) \nabla \psi({\bf r}) \cdot \nabla \frac{\delta F_{\rm total}}{\delta \phi_{\rm m}} + o(H^{2}).
\end{equation}   
In order to obtain a closed equation for $\psi({\bf r})$, 
we rewrite $\delta F_{\rm total}/ \delta \phi_{\rm m}({\bf r})$ in eq.(\ref{13_7}) in terms of $\delta F_{\rm total}/ \delta \psi({\bf r})$.  
For this purpose, we use the following formula:
\begin{equation}
\label{13_8}
\frac{\delta F_{\rm total}(\{ \psi \})  }{\delta \psi({\bf r})} 
= \int \frac{\delta F_{\rm total}(\{ \psi \} )}{ \delta \phi_{\rm m}({\bf r}')} \frac{\delta \phi_{\rm m}({\bf r}')}{\delta \psi ({\bf r})} d{\bf r}'. 
\end{equation}
With use of the first line of eq.(\ref{13_6}), we obtain the functional derivative of $\phi_{\rm m}({\bf r})$ with respect to $\psi({\bf r})$ as follows:
\begin{eqnarray}
\label{13_9}
\frac{\delta \phi_{\rm m}({\bf r}')}{\delta \psi({\bf r})} &=& 2 \epsilon^{2} \nabla' \delta({\bf r}-{\bf r}') \cdot \nabla' \psi({\bf r}') \nonumber \\
&~& -2\psi({\bf r}') \left( 1 - \psi({\bf r}')^{2} \right) \delta({\bf r} - {\bf r}'). 
\end{eqnarray}
Substituting eq.(\ref{13_9}) into eq.(\ref{13_8}), we obtain
\begin{eqnarray}
\label{13_10}
\frac{\delta F_{\rm total}(\{ \psi \})  }{\delta \psi({\bf r})}
= -2 \epsilon^{2} \nabla \psi ({\bf r}) \cdot \nabla \frac{\delta F_{\rm total} (\{ \psi \}) }{\delta \phi_{\rm m}(\bf r )} \nonumber \\
+ 2 H({\bf r}) \frac{\delta F_{\rm total} ( \{ \psi  \} ) }{\delta \phi_{\rm m}({\bf r})},
\end{eqnarray}
where we used the relation $H({\bf r})= -\psi({\bf r}) + \psi({\bf r})^{3} - \epsilon^{2} \nabla^{2} \psi({\bf r}) $.  
From eq.(\ref{13_10}), we obtain
\begin{eqnarray}
\label{13_11}
\nabla \psi({\bf r}) \cdot \nabla \frac{\delta F_{\rm total}}{\delta \phi_{\rm m}({\bf r})} 
&=& - \frac{1}{2 \epsilon^{2}} \frac{\delta F_{\rm total}}{\delta \psi({\bf r})} + \frac{1}{\epsilon^{2}} H({\bf r}) \frac{\delta F_{\rm total}}{\delta \phi_{\rm m}({\bf r})} \nonumber \\
&=& - \frac{1}{2 \epsilon^{2}} \frac{\delta F_{\rm total}}{\delta \psi({\bf r})} + o(H^{2}).
\end{eqnarray}
Substituting eq.(\ref{13_11}) into eq.(\ref{13_7}), we obtain the solution of eq.(\ref{psi_time_1}) that is correct up to the 1st order in $H$ as
\begin{equation}
\frac{\partial \psi({\bf r})}{\partial t} = - \frac{L({\bf r})}{2 \epsilon^{2}} \frac{\delta F_{\rm total}}{\delta \psi({\bf r})} + o(H^{2}).
\end{equation}
This equals to eq.(\ref{noncons}).

\end{document}